\documentclass[twocolumn,showpacs,aps,prl]{revtex4}

\usepackage{graphicx}%
\usepackage{dcolumn}
\usepackage{amsmath}

\begin{document}

\title{Information loss in an optimal maximum likelihood decoding}

\author{In\'es Samengo}

\email{samengo@cab.cnea.gov.ar}

\affiliation{Centro At\'omico Bariloche and Instituto Balseiro \\
(8400) San Carlos de Bariloche, R\'{\i}o Negro, Argentina}


\begin{abstract}
The mutual information between a set of stimuli and the elicited
neural responses is compared to the corresponding decoded
information. The decoding procedure is presented as an artificial
distortion of the joint probabilities between stimuli and
responses. The information loss is quantified. Whenever the
probabilities are only slightly distorted, the information loss
is shown to be quadratic in the distortion
\end{abstract}

\maketitle

Understanding the way external stimuli are represented at the
neuronal level is one central challenge in neuroscience. An
experimental approach to this end (Optican and Richmond 1987,
Eskandar {\it et al.} 1992, Tov\'ee {\it et al.} 1993, Kjaer {\it
et al.} 1994, Heller {\it et al.} 1995, Rolls {\it et al.} 1996,
Treves {\it et al.} 1996, Rolls {\it et al.} 1997, Treves 1997,
Rolls and Treves 1998, Rolls {\it et al.} 1998) consists in
choosing a particular set of stimuli $s \in {\cal S}$ which can be
controlled by the experimentalist, and exposing these stimuli to
a subject whose neural activity is being registered. The set of
neural responses $r \in {\cal R}$ is then defined as the whole
collection of recorded events. It is up to the researcher to
decide which entities in the recorded signal are considered as
events $r$. For example, $r$ can be defined as the firing rate in
a fixed time window, or as the time difference between two
consecutive spikes, or the $k$ first principal components of the
time variation of the recorded potentials in a given interval,
and so forth.

Once the stimulus set ${\cal S}$ and the response set ${\cal R}$
have been settled, the joint probabilities $P(r, s)$ may be
estimated from the experimental data. This is usually done by
measuring the frequency of the joint occurrence of stimulus $s$
and response $r$, for all $s \in {\cal S}$ and $r \in {\cal R}$.
The mutual information between stimuli and responses reads
(Shannon 1948)
\begin{equation}
I = \sum _s \sum _r P(r, s) \ \log _2 \left[\frac{P(r, s)}{P(r) P(s)}
\right],
\label{inf}
\end{equation}
where
\begin{eqnarray}
P(r) &=& \sum _s P(r, s) \label{ueja1} \\
P(s) &=& \sum _r P(r, s). \label{ueja2}
\end{eqnarray}
The mutual information quantifies how much can be learned about
the identity of the stimulus shown just by looking at the
responses. Accordingly, and since $I$ is symmetrical in $r$ and
$s$, its value is also a measure of the amount of information
that the stimuli give about the responses. From a theoretical
point of view, $I$ is the most appealing quantity characterizing
the degree of correlation between stimuli and responses that can
be defined. This stems from the fact that $I$ is the only additive
functional of $P(r, s)$ ranging from zero (for uncorrelated
variables) up to the entropy of stimuli or responses (for a
deterministic one to one mapping) (Fano 1961, Cover and Thomas
1991).

However, even if formally sound, the mutual information has a
severe drawback when dealing with experimental data. Many times,
and specifically when analyzing data of multi-unit recordings,
the response set ${\cal R}$ is quite large, its size increasing
exponentially with the number of neurons sampled. Therefore, the
estimation of $P(r, s)$ from the experimental frequencies may be
far from accurate, specially when recording from the vertebrate
cortex, where there are long time scales in the variability and
statistical structure of the responses. The mutual information
$I$, being a non linear function of the joint probabilities, is
extremely sensitive to the errors that may be involved in their
measured values. As derived in Treves and Panzeri (1995), Panzeri
and Treves (1996) and Golomb {\it et al.} (1997), the mean error
in calculating $I$ from the frequency table of events $r$ and $s$
is linear in the size of the response set. This analytical result
has been obtained under the assumption that different responses
behave independently. Although there are situations where such a
condition does not hold (Victor and Purpura, 1997) it is widely
accepted that the bias grows rapidly with the size of the
response set.

Therefore, a common practice when dealing with large response
sets is to calculate the mutual information not between ${\cal
S}$ and ${\cal R}$, but between the stimuli and another set
${\cal T}$ each of whose elements $t$ is a function of the true
response $r$, that is, $t = t(r)$ (Treves 1997, Rolls and Treves
1998). It is easy to show that if the mapping between $r$ and $t$
is one to one, then the mutual information between ${\cal S}$ and
${\cal R}$ is the same as the one between ${\cal S}$ and ${\cal
T}$. However, for one to one mappings, the number of elements in
${\cal T}$ is the same as in ${\cal R}$. A wiser procedure is to
choose a set ${\cal T}$ that is large enough not to lose the
relevant information, but sufficiently small as to avoid
significant limited sampling errors. One possibility is to perform
a decoding procedure (Gochim {\it et al.} 1994, Rolls {\it et al.}
1996, Victor and Purpura 1996, Rolls and Treves 1998). In this
case, ${\cal T}$ is taken to coincide with ${\cal S}$. To make
this correspondence explicit, the set ${\cal T}$ will be denoted
by ${\cal S}'$ and its elements $t$ by $s'$. Each $s'$ in ${\cal
S}'$ is taken to be a function of $r$, and is called the {\it
predicted stimulus} of response $r$. As stated in Panzeri {\it et
al.} (1999), this choice for ${\cal T}$ is the smallest that
could potentially preserve the information of the identity of the
stimulus. The data processing theorem (Cover and Thomas, 1991)
states that since $s'$ is a function of $r$ alone, and not of the
true stimulus $s$ eliciting response $r$, the information about
the real stimulus can only be lost and not created by the
transformation from $r \to s'$. Therefore, the true information
$I$ is always at least as large as the decoded information $I_D$,
the latter being the mutual information between ${\cal S}$ and
${\cal S}'$\footnote{It should be kept in mind, however, that
when $I_D$ is calculated from actual recordings, its value is
typically overestimated, because of limited sampling. Therefore,
when dealing with real data sets, one may eventually obtain a
value for $I_D$ that surpasses the true mutual information $I$.
Nevertheless, whenever the number of elements in ${\cal S}'$ is
significantly smaller than the number of responses $r$, the
sampling bias in $I_D$ will be bound by the one obtained in the
estimation of $I$.}. In order to have $I$ and $I_D$ as close as
possible, it is of course necessary to choose the best $s'$ for
every $r$. The procedure consists in identifying which of the
stimuli was most probably shown, for every elicited response. The
conditional probability of having shown stimulus $s$ given that
the response was $r$ reads
\begin{equation}
P(s | r) = \frac{P(r, s)}{P(r)}.
\label{bayes}
\end{equation}

Therefore, the stimulus that has most likely elicited response $r$ is
\begin{equation}
s'(r) = \max _s P(s | r) = \max _s P(r, s).
\label{sprim}
\end{equation}
By means of Eq. (\ref{sprim}), a mapping $r \to s'$ is
established: each response has its associated maximum likelihood
stimulus. Equation (\ref{bayes}) provides the only definition of
$P(s | r)$ that strictly follows Bayes' rule, so in this case,
the decoding is called {\it optimal}. There are other alternative
ways of defining $P(s | r)$ (Georgopoulos {\it et al.} 1986,
Wilson and McNaughton 1993, Seung and Sompolinsky 1993, Rolls
{\it et al.} 1996) some of which have the appealing property of
being simple enough to be plausibly carried out by downstream
neurons themselves. The purpose of this letter, however, is to
quantify how much information is lost when passing from $r$ to
$s'$ using an optimal maximum likelihood decoding procedure.

In general, there are several $r$ associated with a given $s'$.
One may therefore partition the response space ${\cal R}$ in
separate classes ${\cal C}(s) = \{r / s'(r) = s \}$, one class
for every stimulus. The number of responses in class $s'$ is
$N_{s'}$. Of course, some classes may be empty. Here, the
assumption is made that each $r$ belongs to one and only class
(that is, Eq. (\ref{sprim}) has a unique solution).

The joint probability of showing stimulus $s$
and decoding stimulus $s'(r)$ reads
\begin{equation}
P(s', s) = \sum _{r \in {\cal C}(s')} P(r, s),
\label{psprims}
\end{equation}
and the overall probability of decoding $s'$,
\begin{equation}
P(s') = \sum _s P(s', s) = \sum _{r \in {\cal C}(s')} P(r).
\label{psprim}
\end{equation}
Clearly, with these definitions the decoded information
\begin{equation}
I_D = \sum _s \sum _{s'} P(s', s) \ \log _2 \left[ \frac{P(s', s)}{P(s')
P(s)} \right]
\label{id}
\end{equation}
may be calculated, and has, in fact, been used in several
experimental analyses (Rolls {\it et al.} 1996, Treves 1997, Rolls
and Treves 1998, Panzeri {\it et al.} 1999). However, up to date,
no rigorous relationship between $I$ and $I_D$ has been
established. The derivation of such a relationship is the main
purpose here.

When performing a decoding procedure, $r$ is replaced by $s'$.
Such a mapping allows the calculation of $P(s', s)$, after which
any additional structure, which may eventually have been present
in $P(r, s)$, is neglected. For example, if two responses $r_1$
and $r_2$ encode the same stimulus $s'$ it becomes irrelevant
whether, for a given $s$, $P(r_1, s)$ is much bigger that $P(r_2,
s)$ or, on the contrary, $P(r_1, s) \approx P(r_2, s)$. The only
thing that matters is the value of the sum of the two: their
global contribution to $P(s', s)$. As a consequence, it seems
natural to consider the detailed variation of $P(r, s)$ {\it
within} each class, when estimating the information lost in the
decoding.

In this spirit, and aiming at quantizing such a loss of
information, $P(r, s)$ is written as
\begin{equation}
P(r, s) = \frac{P\left[s'(r), s \right]}{N_{s'(r)}} + \Delta(r, s),
\label{sep1}
\end{equation}
where $\Delta(r, s) = P(r, s) - P[s'(r), s]/N_{s'(r)}$. Thus, the
joint probability $P(r, s)$, which in principle may have quite a
complicate shape in ${\cal R}$ space, is separated into two terms. The
first one is flat inside every single class ${\cal C}(s')$, and
the second is whatever needed to re-sum $P(r, s)$. It should be
noticed that
\begin{equation}
\sum _{r \in {\cal C}(s')} \Delta(r, s) = 0,
\label{sum1}
\end{equation}
for all $s$. Summing Eq. (\ref{sep1}) in $s$,
\begin{equation}
P(r)= \frac{P\left[s'(r)\right]}{N_{s'(r)}} + \Delta(r),
\label{sep2}
\end{equation}
where
\begin{equation}
\Delta(r) = \sum _s \Delta(r, s),
\label{overd}
\end{equation}
and
\begin{equation}
\sum _{r \in {\cal C}(s')} \Delta(r) = 0.
\label{sum2}
\end{equation}
Replacing Eqs. (\ref{sep1}) and (\ref{sep2}) in the mutual
information (\ref{inf}), one arrives at
\begin{equation}
I = I_D + \sum_r \sum_s P(r, s) \log_2 \left[ \frac{P(r, s)}{Q(r,
s)} \right], \label{descomp}
\end{equation}
where
\begin{equation}
Q(r, s) = \frac{P[s'(r), s]}{N_{s'}} + \Delta(r) \frac{P[s'(r),
s]} {P(s')}
\end{equation}
is a properly defined distribution, since it can be shown to be
normalized and non-negative. The term in the right of Eq.
(\ref{descomp}) is the Kullback-Leibler divergence (Kullback 1968)
between the distributions $P$ and $Q$, which is guaranteed to be
non negative. This confirms the intuitive result $I_D \le I$, the
equality being only valid when
\begin{equation}
\Delta(r) P[s'(r), s] =\Delta(r, s) P[s'(r)], \label{condicion}
\end{equation}
for all $r$ and $s$.

Equation (\ref{descomp}) states the quantitative difference
between the full and the decoded information, and is the main
result of this letter. The amount of lost information is
therefore equal to the informational distance between the original
probability distribution $P(r, s)$ and a new function $Q(r, s)$.
It can be easily verified that
\begin{equation}
I_D = \sum_s \sum_r Q(r, s) \log_2 \left[ \frac{Q(r, s)}{Q(r)
Q(s)} \right], \label{poing}
\end{equation}
where
\begin{eqnarray}
Q(r) &=& \sum_s Q(r, s) = P(s), \nonumber \\
Q(s) &=& \sum_r Q(r, s) = P(r). \label{chuik}
\end{eqnarray}
Therefore, the decoded information can be interpreted as a full
mutual information between the stimuli and the responses, but with
a distorted probability distribution $Q(r, s)$. In this context,
the difference $I - I_D$ is no more than the distance between the
true distribution $P(r, s)$ and the distorted one $Q(r, s)$.

When is Eq. (\ref{condicion}) fulfilled? Surely, if there is at
most one response in each class, $\Delta$ is always zero, and $I
= I_D$. Also, if $P(r, s)$ is already flat in each class, there
is no information loss. However, if $P(r, s)$ is not flat inside
every class, but obeys the condition $P(r, s) = P_{s'}(r) P(s',
s)$ for a suitable $P(s', s)$ and some function $P_{s'}(r)$ that
sums up to unity within ${\cal C}(s')$, one can easily show that
Eq. (\ref{condicion}) holds. Just notice that this case implies
that if $r_1$ and $r_2$ belong to ${\cal C}(s')$, then $P(r_1, s)
/ P(r_2, s)$ is independent of $r$, for all $s$. In other words,
within each class ${\cal C}(s')$, the different functions $P(r |
s)$ obtained by varying $s$ differ from one another by a
multiplicative constant. These conditions coincide with the ones
given by Panzeri {\it et al.} (1999) for having an exact
decoding, within the short time limit. However, in the present
derivation there are no assumptions about the interval in which
responses are measured. Therefore, the decoding being exact
whenever Eq. (\ref{condicion}) is fulfilled is not a consequence
of the short time limit carried out by Panzeri {\it et al.}
(1999), but rather, a general property of the maximum likelihood
decoding.

Next, by making a second order Taylor expansion of Eq.
(\ref{descomp}) in the distorsions $\Delta(r, s)$ and $\Delta(r)$
one may show that
\begin{equation}
I = I_D + \sum _s \sum _{s'} P(s', s) \frac{E(s', s)}{2 \ln 2} +
{\cal O}(\Delta^2), \label{fin}
\end{equation}
where
\begin{equation}
E(s', s) = \frac{1}{N_{s'}} \sum _{r \in {\cal C}(s')}
\left[\left(\frac{\Delta(r, s)}{P(s', s) / N_{s'}} \right)^2 -
\left(\frac{\Delta(r)}{P(s')/N_{s'}} \right)^2  \right]
\label{estax}
\end{equation}
Therefore, in the small $\Delta$ limit, the difference between
$I$ and $I_D$ is quadratic in the distortions $\Delta(r, s)$ and
$\Delta(r)$. This means that if in a given situation these
quantities are guaranteed to be small, then the decoded
information will be a good estimate of the full information.
Equation (\ref{estax}) is equivalent to
\begin{equation}
E(s', s) =  \left \langle \left(\frac{P(r, s)}{P(s', s) / N_{s'}}
\right)^2 - \left(\frac{P(r)}{P(s')/N_{s'}} \right)^2  \right
\rangle _{{\cal C}(s')}, \label{error}
\end{equation}
where
\[
\langle f(r) \rangle_{{\cal C}(s')} = \frac{1}{N(s')} \ \sum_{r
\in {\cal C}(s')} f(r).
\]
As a consequence, the relevant parameter in determining the size
of $E(s', s)$ is given by the mean value---within ${\cal
C}(s')$---of a function that essentially measures how different
are the true probability distributions $P(r, s)$ and $P(r)$, from
their flattened versions $P(s', s) / N_{s'}$ and $P(s')/N_{s'}$.

To summarize, this letter presents the maximum likelihood decoding
as an artificial---but useful---distortion of the distribution
$P(r, s)$ within each class ${\cal C}(s')$. The decoded
information is shown to be also a mutual information, the latter
calculated with the distorted probability distribution. The
difference between $I$ and $I_D$ is the Kullbach-Leibler distance
between the true and distorted distributions. As such, it is
always non negative, and it is easy to identify the conditions for
the equality between the two information measures. Finally, for
small distortions $\Delta$, the amount of lost information is
expressed as a quadratic function in $\Delta$. In short, the aim
of the work is to present a formal way of quantizing the effect
of an optimal maximum likelihood decoding.

It should be kept in mind that in real situations, where only a
limited amount of data is available, the estimation of $P(r | s)$
may well involve a careful analysis in itself. Some kind of
assumption (as for example, a Gaussian shaped response
variability) is usually required. The validity of the assumptions
made depend on the particular data at hand.  An inadequate choice
for $P(r | s)$ may of course lead to a distorted value of $I$,
and in fact, the bias may be in either direction. If the choice
of $P(r | s)$ does not even allow the correct identification of
the maximum likelihood stimulus (see Eq. (\ref{sprim})), then the
calculated value of $I_D$ will also be distorted. The purpose of
this letter, however, is to quantify how much information is lost
when passing from $r$ to $s'(r)$. No attempt has been made to
quantify $I$ or $I_D$, for different estimations of $P(r |s)$.

Sometimes, $P(s', s)$ is defined in terms of $P(r, s)$ without
actually decoding the stimulus to be associated to each response.
For example, $P(s', s)$ can be introduced as $\sum_r P(r, s')
P(r, s) / P^2(r)$ (Treves, 1997). This approach, although
formally sound, is not based in a $r \rightarrow s'$ mapping, and
does not allow a partition of ${\cal R}$ into classes. It is
therefore is not directly related to the analysis presented here.
However, there might be analogous derivations where one may get
to quantify the information loss also in this case.

\section*{Acknowledgements}

I thank Bill Bialek, Anna Montagnini and Alessandro Treves for
very useful discussions. This work has been partially supported
with a grant of Proff. Treves, of the Human Frontier Science
Program, number RG 01101998B.

\section*{References}

\begin{itemize}

\item[-] Bialek, W., Rieke, F., de Ruyter van Steveninck, R. R.,
\& Warland, D. (1991). Reading a neural code. {\it Science}, {\bf
252}, 1854 - 1857.

\item[-] Cover, M. T., \& Thomas, J. A., (1991). {\it Elements
of Information Theory}. New York: Wiley.

\item[-] Eskandar, E. N., Richmond, B. J., \& Optican, L., M.
(1992). Role of inferior temporal neurons in visual memory. I.
Temporal encoding of information about visual images, recalled
images, and behavioural context. {\it J. Neurophysiol.}, {\bf
68}, 1277 - 1295.

\item[-] Fano, R. M. (1961) {\it Transmission of Information: A
Statistical Theory of Communications}. New York: MIT.

\item[-] Georgopoulos, A. P., Schwartz, A., \& Kettner, R. E.
(1986). Neural population coding of movement direction. {\it
Science}, {\bf 233}, 1416 - 1419.

\item[-] Gochin, P. M., Colombo, M., Dorfman, G. A., Gerstein, G.
L., \& Gross, C. G. (1994). Neural ensemble encoding in inferior
temporal cortex. {\it J. Neurophysiol}, {\bf 71}, 2325 - 2337.

\item[-] Golomb, D., Hertz, J., Panzeri, S., Treves, A., \& Richmond,
B. (1997). How well can we estimate the information carried in
neuronal responses from limited samples? {\it Neural Comp.}, {\bf
9}, 649 - 655.

\item[-] Heller, J., Hertz, J. A., Kjaer, T. W., \& Richmond, B. J.
(1995). Information flow and temporal coding in primate pattern
vision. {\it J. Comput. Neurosci.}, {\bf 2}, 175 - 193.

\item[-] Kjaer, T. W., Hertz, J. A., \& Richmond, B. J. (1994).
Decoding cortical neuronal signals: networks models, information
estimation and spatial tuning. {\it J. Comput. Neurosci.}, {\bf
1}, 109 - 139.

\item[-] Kullback, S., (1968). {\it Information theory and statistics}.
New York: Dover.

\item[-] Optican, L. M., \& Richmond, B. J. (1987). Temporal encoding
of two dimensional patterns by single units in primate inferior
temporal cortex: III Information theoretic analysis. {\it J.
Neurophysiol.}, {\bf 57}, 162 - 178.

\item[-] Panzeri, S., \& Treves, A. (1996). Analytical estimates of
limited sampling biases in different information measures. {\it
Network}, {\bf 7}, 87 - 107.

\item[-] Panzeri, S., Treves, A., Schultz, S., \& Rolls, E. T.
(1999). On decoding the responses of a population of neurons from
short time windows. {\it Neural Comput.}, {\bf 11}, 1553 - 1577.

\item[-] Rieke, R., Warland, D., de Ruyter van Steveninck, R. R.,
\& Bialek W., (1996). {\it Spikes: Exporing the Neural Code}.
Cambridge: MIT Press.

\item[-] Rolls, E. T., Critchley, H. D., \& Treves, A. (1996).
Representation of Olfactory Information in the Primate
Orbitofronal Cortex. {\it J. Neurophysiol.}, {\bf 75}, (5), 1982 -
1996.

\item[-] Rolls, E. T., Treves, A., \& Tov\'ee, M. J. (1997).
The representational capacity of the distributed encoding of
information provided by populations of neurons in primate
temporal visual area. {\it Exp. Brain. Res.}, {\bf 114}, 149 -
162.

\item[-] Rolls, E. T., \& Treves, A. (1998). {\it Neural Networks and
Brain Function}. Oxford: Oxford University Press.

\item[-] Rolls, E. T., Treves, A., Robertson, R. G., Georges-Francois, P.,
\& Panzeri, S. (1998). Information About Spatial View in an
Ensemble of Primate Hippocampal Cells. {\it J. Neurophysiol.},
{\bf 79}, 1797 - 1813.

\item[-] de Ruyter van Steveninck, R. R., \& Laughlin, S. B.
(1996). The rates of information transfer at graded-potential
synapses. {\it Nature}, {\bf 379}, 642 - 645.

\item[-] Seung, H. S., \& Sompolinsky, H. (1993). Simple models for
reading neural population codes. {\it Proc. Nac. Ac. Sci. USA},
{\bf 90}, 10749 - 10753.

\item[-] Shannon, C. E. (1948). {\it AT\&T Bell Laboratories Technical
Jounal} ,{\bf 27}, 379 - 423.

\item[-] Tov\'ee, M. J., Rolls, E. T., Treves, A. \& Bellis, R. J.
(1993). Information encoding and the responses of single neurons
in the primate temporal visual cortex. {\it J. Neurophysiol.},
{\bf 70}, 640 - 654.

\item[-] Treves, A., \& Panzeri, S. (1995). The upward bias in measures
of information derived from limited data samples. {\it Neural
Comp.}, {\bf 7}, 399 - 407.

\item[-] Treves, A., Skaggs, W. E., \& Barnes, C. A. (1996).
How much of the hippocampus can be explained by functional
constraints? {\it Hippocampus}, {\bf 6}, 666 - 674.

\item[-] Treves, A. (1997). On the perceptual structure of face
space. {\it BioSyst.}, {\bf 40}, 189 - 196.

\item[-] Victor, J. D., \& Purpura, K. P. (1996). Nature and precision
of temporal coding in visual cortex: a metric space analysis. {\it
J. Neurophysiol.}, {\bf 76}, 1310 - 1326.

\item[-] Victor, J. D., \& Purpura, K. P. (1997). Metric-space analysis
of spike trains: theory, algorithms and application. {\it Network}
{\bf 8} 127 - 164

\item[-] Wilson, M. A., \& McNaughton, B. L. (1993). Dynamics of the
Hippocampal Ensemble Code for Space. {\it Science}, {\bf 261},
1055 - 1058.

\end{itemize}

\end{document}